\documentclass[aip,jcp,reprint]{revtex4-1}
\usepackage{graphicx}
\usepackage{amsmath}
\usepackage{color}

\newcommand{\bnu}{{\boldsymbol\nu}}
\newcommand{\bphi}{{\boldsymbol\phi}}
\newcommand{\bxi}{{\boldsymbol\xi}}
\newcommand{\bpsi}{{\boldsymbol\psi}}

\draft % marks overfull lines with a black rule on the right

\begin{document}

%\preprint{}

\title{Diffusion approximations to the chemical master equation only have a consistent stochastic thermodynamics at chemical equilibrium} %Title of paper



\author{Jordan M. Horowitz}
\email[]{jordan.horowitz@umb.edu}
%\thanks{}
\affiliation{Department of Physics, University of Massachusetts at Boston, Boston, MA 02125, USA}

\date{\today}

\begin{abstract}
The stochastic thermodynamics of a dilute, well-stirred mixture of chemically-reacting species is built on the stochastic trajectories of reaction events obtained from the Chemical Master Equation.
However, when the molecular populations are large, the discrete Chemical Master Equation can be approximated with a continuous diffusion process, like the Chemical Langevin Equation or Low Noise Approximation.
In this paper, we investigate to what extent these diffusion approximations inherit the stochastic thermodynamics of the Chemical Master Equation.
We find that a stochastic-thermodynamic description is only valid at a detailed-balanced, equilibrium steady state.
Away from equilibrium, where there is no consistent stochastic thermodynamics, we show that one can still use the diffusive solutions to approximate the underlying thermodynamics of the Chemical Master Equation.
\end{abstract}

\pacs{82.20.-w, 82.60.-s, 05.70.Ln, 05.40.-a}% insert suggested PACS numbers in braces on next line

\maketitle %\maketitle must follow title, authors abstract and \pacs

\section{Introduction}\label{sec:intro}

In a dilute, well-stirred mixture of chemically-reacting species, the Chemical Master Equation (CME)  describes the stochastic dynamics of the molecular populations~\cite{VanKampen}.
Despite the accuracy of the CME, it is often difficult to use; analytic solutions are rare and its simulation is often challenging~\cite{VanKampen,Gardiner,Gillespie2007}.
This has lead to a number of systematic expansions and approximations.
In particular, when the molecular populations are large, the discrete nature of the chemical reactions smooths out, giving rise to an approximate continuous diffusion process~\cite{Gardiner}.
Two of the most influential such approaches are Gillespie's Chemical Langevin Equation~\cite{Gillespie2000} (CLE) and van Kampen's Low Noise Approximation~\cite{VanKampen} (LNA), or System Size Expansion.
These techniques are significantly more tractable than the CME, both for analytical as well as computational calculations.

Insight into the structure and function of chemical reaction networks comes not just from studying their population dynamics, but also from their energetics and thermodynamics.
For nonequilibrium fluctuating systems, like those described by the CME, there has emerged a robust theoretical framework that not only treats the average thermodynamic behavior, but also the fluctuations.
This framework, called stochastic thermodynamics, ascribes thermodynamic quantities -- such as heat, work, and entropy -- to individual, fluctuating trajectories~\cite{Sekimoto,Seifert2012,VandenBroeck2015}.
This point of view has been fruitful in understanding fundamental aspects of far-from-equilibrium systems.
In particular, it has aided in the development of a collection of exact far-from-equilibrium equalities known as the fluctuation theorems~\cite{Jarzynski2011,Harris2007}, which have have refined our understanding of thermodynamic irreversibility.

In general, stochastic thermodynamics is composed of two main ingredients that quantify the energy balance and entropy balance along an individual, stochastic trajectory $\gamma$ of the system's  dynamics.
 The first ingredient is an application of  the first law of thermodynamics to every trajectory, relating the change in internal energy $\Delta e[\gamma]$ to the work done on the system $w[\gamma]$ and  the heat released into the surroundings  $q[\gamma]$:
\begin{equation}
\Delta e[\gamma] = w[\gamma]-q[\gamma].
\end{equation}
The second ingredient is a trajectory-dependent total entropy production, which measures the trajectories'  thermodynamic irreversibility.
It is obtained as the log-ratio of the probability to observe a particular trajectory ${\mathcal P}[\gamma]$ to the probability of realizing the time-reversed trajectory ${\mathcal P}[\tilde\gamma]$:
\begin{equation}\label{eq:2law}
\Delta s^{\rm tot}[\gamma]= \Delta s[\gamma]+\Delta s^{\rm e}[\gamma] = \ln\frac{{\mathcal P}[\gamma]}{{\mathcal P}[\tilde\gamma]},
\end{equation}
where we have included the customary splitting into the change in system entropy $\Delta s$ and (environmental) entropy flow $\Delta s^{\rm e}$.
The traditional statement of the second law -- that entropy production is positive -- emerges only on average, $\Delta S^{\rm tot}=\langle \Delta s^{\rm tot}\rangle \ge 0$.

Notice that the entropy production $\Delta s^{\rm tot}$ is obtained solely from the dynamics, making no reference to the energetics.
This disconnect is in stark contrast to macroscopic thermodynamics, where the heat enters in the second law as the entropy flow into the environment~\cite{Callen}.
Without this connection, the second law does not provide bounds on the energy requirements of  thermodynamic processes.
Therefore, a consistent stochastic-thermodynamic description requires that the heat dissipated into the environment along any trajectory have a well-defined entropy increase, that is we require
\begin{equation}
\Delta s^{\rm e}[\gamma]=\beta q[\gamma],
\end{equation}
where $\beta = 1/k_{\rm B}T$ is the inverse temperature of the environment and $k_{B}$ is the Boltzmann constant.
When such a connection exists, we say that the stochastic thermodynamics is consistent~\cite{Seifert2012}.  
For jump processes, consistency requires that the microscopic transition rates verify a local detailed balance relation~\cite{Esposito2010b,VandenBroeck2015}, while for diffusion processes the fluctuation-dissipation theorem imposes consistency~\cite{Seifert2012,Lebowitz1999}.

A consistent stochastic thermodynamics for the CME was laid out by Schmiedl and Seifert~\cite{Schmiedl2007b}, building on earlier studies of biochemical models of enzymes~\cite{Schmiedl2007} and fluctuation theorems for nonequilibrium reactions~\cite{Gaspard2004, Andrieux2004}; but one can trace the seeds of this framework, at least for the average thermodynamic behavior, to Hill's classic text~\cite{Hill}, Schnakenberg's network theory~\cite{Schnakenberg1976}, and Qian's analysis of the average energetics~\cite{Qian2001,Qian2005,Qian2005b}.  
Since the CME has a consistent stochastic thermodynamics, one may wonder if the CME's diffusion approximations inherit that structure.
As a first attempt to answer this question, an entropy production [like Eq.~(\ref{eq:2law})] was constructed for the CLE by Xiao, Hou, and Xin~\cite{Xiao2009} and for the LNA by Tomita and Sano~\cite{Tomita2008}.
Both groups found no concrete connection between the entropy production and energetics.
In this paper, we reanalyze this problem and demonstrate that with a proper identification of the entropy production a consistent thermodynamics only emerges for the CLE and LNA at equilibrium.
In particular, the entropy flows for these diffusion approximations only coincide with the CME entropy flow (and therefore heat flow) at equilibrium.
Away from equilibrium, we show that we can still use the CLE and LNA to approximate the heat flow of the CME, but this approximate heat flow will not be related to the trajectory entropy production of the CLE and LNA as one would want for a consistent stochastic thermodynamics.

The outline is as follows: In Sec.~\ref{sec:dynamics}, we describe the setup and review the dynamics and thermodynamics of the CME, CLE, LNA, and the macroscopic reaction rate equations (RRE).  Then in Sec.~\ref{sec:equilibrium} the entropy flows for all the equations of motion are compared at equilibrium using simple approximations; supporting detailed calculations are presented in the Appendices.  In Sec.~\ref{sec:approx}, diffusive approximations to the CME's entropy flow are developed.   Our results are then illustrated in Sec.~\ref{sec:example} with a linear chemical reaction network for one species, before concluding in Sec.~\ref{sec:conclusion}.
  
 \section{Dynamics and Irreversibility}\label{sec:dynamics}
 
 \subsection{Setup}

 We have in mind a well-stirred mixture of $N$ chemical species with time-dependent, molecular populations ${\bf X}_t = \{X_1(t),\dots, X_N(t)\}$  in a fixed volume $\Omega$ at constant temperature $T=1/\beta$ (in $k_{\rm B}=1$ units, which we assume throughout).
 The molecular populations change randomly in discrete jumps through $M$ reversible reaction channels, denoted as $R_\rho$ and $R_{-\rho}$ ($\rho=1,\dots, M$) for the forward and reverse reactions, respectively.
 The corresponding reaction equations are
  \begin{equation}
 \begin{split}
&{\bf X} \xrightarrow{R_{\rho}} {\bf X}+\boldsymbol\nu^\rho,
\end{split}
\end{equation}
where each element of the vector of stochiometric coefficients $\boldsymbol\nu^\rho = \{\nu_i^\rho\}$ gives the change in species $X_i$ during reaction $R_\rho$.
Accordingly, the stochiometric vectors for a pair of forward and reverse reactions are related by $\boldsymbol\nu^{-\rho} = -\boldsymbol\nu^\rho$.

The stochastic jump dynamics of the above chemical reaction network is predicated on the existence of a collection of propensities $a_\rho({\bf x})$ that give the probability rate for reaction $R_\rho$ to occur in an infinitesimal time interval  given the populations ${\bf x}$~\cite{VanKampen, Gillespie2000}. 
For example, in a unimolecular reaction~\cite{Gillespie2000} the propensity is $a=cx$ with rate constant $c$.
Alternatively, if we describe the dynamics using the molecular concentrations ${\bf Z}_t={\bf X}_t/\Omega$, the propensities have the scaling $a_{\rho}({\bf x}) = \Omega \alpha_{\rho}({\bf z})$ (at least approximately for large $\Omega$)~\cite{Gillespie2000}; in our unimolecular example, $\alpha=c z$.

 \subsection{Chemical Master Equation}
 
 The vector of molecular populations ${\bf X}_t$ is a time-dependent random variable whose stochastic evolution tracks the changing number of molecules due to the discrete, random chemical reactions.
 As such its evolution is a jump-type Markov process with rates $a_\rho({\bf x})$ given by the Chemical Master Equation~\cite{VanKampen,Gillespie2000} for the time-dependent probability $P_t({\bf x)}$ of the populations ${\bf X}_t={\bf x}$,
 \begin{equation}\label{eq:CME}
 \frac{\partial }{\partial t}P_t({\bf x}) =\sum_{\rho=\pm1}^{\pm M}  a_\rho({\bf x}-{\boldsymbol\nu}^\rho)P_t({\bf x}-\boldsymbol\nu^\rho)-a_\rho({\bf x}) P_t({\bf x}).
 \end{equation}
 We assume that the CME has a unique steady-state distribution $P_{\rm ss}$ given as the solution of
 \begin{equation}
 \sum_{\rho=\pm 1}^{\pm M}  a_\rho({\bf x}-{\boldsymbol\nu}^\rho)P_{\rm ss}({\bf x}-\boldsymbol\nu^\rho)-a_\rho({\bf x}) P_{\rm ss}({\bf x})=0.
 \end{equation}
 In the special case where each reaction is individually balanced by its reverse in the steady state,
 \begin{equation}\label{eq:DB}
 a_\rho({\bf x}-{\boldsymbol\nu}^\rho)P_{\rm ss}({\bf x}-\boldsymbol\nu^\rho)=a_{-\rho}({\bf x}) P_{\rm ss}({\bf x}),
 \end{equation}
 we say that detailed balance is satisfied~\cite{VanKampen,Schmiedl2007b} and identify the steady-state distribution as the equilibrium  distribution $P_{\rm ss}({\bf x}) = P_{\rm eq}({\bf x})$.
 Let us note that generically the CME steady-state is not unique, but instead is given by a linear combination of  distributions, each confined to a disconnected subset of the state space~\cite{Anderson2010}.
For these cases, one should view our analysis as applied to each such subset separately.
 
 While Eq.~(\ref{eq:CME}) is the standard expression of the CME, we are interested in the stochastic thermodynamics of individual trajectories.
 To make the trajectory picture explicit, we recast the dynamics encoded in the CME as a stochastic differential equation.
 To this end, we introduce a collection of Poisson increments $dN^\rho_t$, which are independent, random variables that are typically $0$, but randomly flicker to $1$ every time reaction $R_\rho$ occurs, giving us a ``click'' that records
 the random occurrence of each reaction~\cite{Wiseman1996}.  
 The rate of ``clicks'' is specified by the conditional expectation $\langle dN^\rho_t\rangle_{\bf x}=a_\rho({\bf x})dt$.
 In terms of these Poisson increments, the population dynamics can be expressed as the It\=o stochastic differential equation 
 \begin{equation}\label{eq:Poisson}
 d{\bf X}_t =  \sum_{\rho\ge 1}\boldsymbol\nu^\rho \left(dN_t^\rho -  dN_t^{-\rho}\right).
  \end{equation}
  ${\bf X}_t$ ticks up by $\boldsymbol\nu^\rho$ every time $R_\rho$ occurs ($dN^\rho_t=1$).
  
 The stochastic description of the reaction dynamics offers an elegant formulation of the thermodynamics in terms of the stochastic increments for the energy and entropy flows~\cite{Horowitz2012}.
  Let us begin with the energy.
 Every time reaction $R_\rho$ occurs, energy is exchanged with the surroundings.
 This energy could take various forms, like thermal or chemical, but as it is exchanged with the surroundings, we identify it as heat $q_\rho({\bf x})$.
  Thus, every time $R_\rho$ occurs, we add $q_\rho$ to our energy accounting, and every time $R_{-\rho}$ occurs we count $q_{-\rho}({\bf x})=-q_\rho({\bf x}-\bnu^\rho)$.
  This coupling of heat fluctuations with chemical reactions suggests that the heat flow along a trajectory changes discretely with stochastic increment~\cite{Schmiedl2007b}
   \begin{equation}\label{eq:CMEheat}
 dq_t = \sum_{\rho\ge 1} q_\rho({\bf X}_t)dN^\rho_t + q_{-\rho}({\bf X}_t) dN^{-\rho}_t.
 \end{equation}
 
 To address the entropy balance, we start by recalling that the system entropy is composed of two pieces.
The first is the internal entropy $s^{\rm int}({\bf x})$, which accounts for the entropy associated to the equilibrated internal degrees of freedom, such as position and momentum.
For example, a single species modeled as a noninteracting gas would have an internal entropy $s^{\rm int}(X) = Xs_1-\ln(X!)$, where $s_1$ is the equilibrium entropy of a single molecule confined to a vessel of volume $\Omega$.
The $\ln(X!)$ term arises due to the indistinguishability of the molecules, which we cannot approximate using Stirling's approximation, as the particle number may be small.
The second contribution is a Shannon-like information entropy for the out-of-equilibrium molecular populations, so that the system entropy reads~\cite{Esposito2012c, Schmiedl2007b}
 \begin{equation}
 \begin{split}
 S&=-\sum_{\bf x}P_t({\bf x})\ln P_t({\bf x}) +\sum_{\bf x} P_t({\bf x})s^{\rm int}({\bf x})\\
 & = \sum_{\bf x} P_t({\bf x})s_t({\bf x})
 \end{split}
 \end{equation}
Now, like the heat flow there is entropy flow into the environment during every reaction.
The first source of entropy flow compensates the change in internal entropy during a reaction, $\Delta s^{\rm int}_\rho({\bf x}) = s^{\rm int}({\bf x}+\bnu^\rho) - s^{\rm int}({\bf x})$.
 The second source is dynamic,
\begin{equation}\label{eq:CMEdynEnt}
\sigma^{\rm CME}_\rho({\bf x}) = \ln\frac{a_\rho({\bf x})}{a_{-\rho}({\bf x}+\boldsymbol\nu^\rho)},
\end{equation}
 which makes explicit that the entropy flow measures how different the likelihood of a  reaction is from its reverse, and in this sense quantifies the irreversibility.
 Thus, the stochastic increment for the entropy flow is~\cite{Schmiedl2007b,Esposito2012c}
 \begin{equation}\label{eq:CMEent}
 \begin{split}
 ds^{\rm e,CME}_t = \sum_{\rho\ge 1}&\left[\sigma^{\rm CME}_\rho({\bf X}_t)-\Delta s^{\rm int}_\rho({\bf X}_t)\right]dN^\rho_t \\
 &+\left[\sigma^{\rm CME}_{-\rho}({\bf X}_t)-\Delta s^{\rm int}_{-\rho}({\bf X}_t)\right] dN^{-\rho}_t.
 \end{split}
 \end{equation}
 Summing the stochastic increment of the system entropy $ds_t$ and the entropy flow $ds_t^{\rm e,CME}$ recovers the entropy production $ds^{\rm tot}_t=ds_t+ds^{\rm e,CME}_t$.

For the stochastic-thermodynamic description to be consistent, the first and second laws need to be related. 
 In particular, the entropy flow should be proportional to the heat along each trajectory: $ds^{\rm e}_t=\beta dq_t$.
 For the CME, this connection holds because the propensities $a_\rho$  are taken to verify the local detailed balance relation~\cite{Seifert2011,Schmiedl2007b,Esposito2012c}
\begin{equation}\label{eq:LDB}
\ln\frac{a_\rho({\bf x})}{a_{-\rho}({\bf x}+\boldsymbol\nu^\rho)}=-\beta\Delta \Phi_{\rho}({\bf x})=\beta q_\rho({\bf x}) + \Delta s^{\rm int}_\rho(\bf {x}),
 \end{equation}
where $\Delta \Phi_{\rho}$ is the change in the populations' grand potential $\Phi_\rho({\bf x})=e({\bf x})-\mu_\rho n({\bf x})-Ts^{\rm int}({\bf x})$  relative to the reservoir mediating reaction $R_\rho$, specifed by the energy $e$, particle number $n$ and the reservoir's chemical potential $\mu_\rho$ and temperature $T$.
The heat $q_\rho=-\Delta e({\bf x})+\mu_\rho \Delta n({\bf x})$ is then due to the change in energy of the molecules $\Delta e$ less the chemical work done by the reservoir $\mu_\rho \Delta n$.

An important instance of local detailed balance occurs with mass action kinetics~\cite{Schmiedl2007b,VanKampen}.
To see this most clearly, let us discuss an illustrative example. 
Consider the addition/subtraction of two molecules of species $A$ into/from the reaction volume due to a chemostat with fixed chemical potential $\mu$, that is, $2A\leftrightarrow \emptyset$.
Let us denote the propensity for adding the molecules as $a_+(A) = k _+e^{2\beta\mu}$ and the removal as $a_-(A) = k_- A(A-1)$.
Consistency with detailed balance then requires that the ratio of the rate constants be $\ln k_+/k_-= -2 \beta f_A$,  where $f_A$ is the free energy of one particle~\cite{Hill,VanKampen,Seifert2011}.
In which case, we have 
\begin{align} \nonumber
\ln\frac{a_+(A)}{a_-(A+2)}&=-2\beta f_A+2\beta\mu-\ln [A(A-1)] \\ \nonumber
&= 2\beta(-e_A+\mu)+(2s_A-\ln[A(A-1)]) \\
& = \beta q_A+\Delta s^{\rm int}(A),
\end{align}
where we have split the single-particle free energy as $f_A=e_A-Ts_A$, identified the change in internal entropy to add two non-interacting molecules to the reaction volume with $\Delta s^{\rm int}(A)=2s_A-\ln[A(A-1)]$, and singled out the heat $q_A=-2(e_A-\mu)$ as the heat flow into the thermal environment $-2e_A$ less the chemical work $2\mu$.
  
Finally, notice that only the dynamic contribution to the entropy production,
\begin{equation}\label{eq:CMEentDyn}
d\sigma^{\rm CME}_t = \sum_{\rho\ge 1}\sigma^{\rm CME}_\rho({\bf X}_t)dN^\rho_t +\sigma^{\rm CME}_{-\rho}({\bf X}_t) dN^{-\rho}_t,
\end{equation}
 is sensitive to the equations of motion: both the system entropy and internal entropy $s^{\rm int}$ are only functions of the molecular populations ${\bf X}_t$.
Thus, only the expression for $\sigma$ will change as we vary the equations of motion.
With this in mind, we will focus on just the dynamic contributions to the entropy flow from now on. 
  
 \subsection{Chemical Langevin Equation}\label{sec:CLEdyn}
 
 Gillespie has argued that there is a regime where the discrete dynamics of the CME can be  approximated by a continuous Langevin equation~\cite{Gillespie2000}.
 Typically, this regime is identified as the large population limit, but Gillespie's analysis shows that the key requirement is the existence of a coarse-grained time scale over which the Poissonian reaction dynamics can be approximated as Gaussian.
 We begin this section with a review  of Gillespie's argument, since a clear understanding of the approximation is vital to characterizing the resulting thermodynamics.
 
Consider a time-interval $\Delta t$ chosen to be long enough that many reactions occur ($a_\rho \Delta t \gg 1$), but short enough that the propensities for each reaction are approximately constant ($a_\rho({\bf X}_{t+\Delta t}) \approx a_\rho({\bf X}_t)$).
 The existence of a suitable $\Delta t$ is dependent on the specifics of the problem, but is likely satisfied for large populations.
In particular, the propensities for mass action kinetics are proportional to the populations, and as such, vary very little during individual reactions: $a_{\rho}({\bf X}_t+\bnu^\rho)\approx a_{\rho}({\bf X}_t)$ for ${\bf X}_t\gg1$.
Thus, many reactions can occur without appreciably altering the reaction rates.
 
 Over the course of $\Delta t$, the change in the molecular populations $\Delta {\bf X}_t$ is obtained by integrating Eq.~(\ref{eq:Poisson}):
 \begin{equation}\label{eq:PoisDisc}
 \Delta{\bf X}_t = \sum_{\rho\ge1} \boldsymbol\nu_\rho (\Delta N^\rho_t - \Delta N^{-\rho}_t),
 \end{equation}
 where $\Delta N^\rho_t = \int_t^{t+\Delta t} dN_s^\rho$ is the Poisson-distributed random number of $R_\rho$ reactions during $\Delta t$.
Now, since the number of reactions is very large and the rate $a_{\rho}$ is approximately constant, we can use the central limit theorem to approximate $\Delta N^\rho_t$ as a Gaussian random variable with mean and variance equal to $a_\rho({\bf X}_t)$:
 \begin{equation}\label{eq:GausApprox}
  \Delta N^\rho_t \approx a_{\rho}({\bf X}_t)\Delta t + \sqrt{a_\rho({\bf X}_t)}\Delta W^\rho_t,
 \end{equation}
 where we have introduced the independent, zero-mean Gaussian random variables $\Delta W^\rho_t$.
Substituting Eq.~(\ref{eq:GausApprox}) into Eq.~(\ref{eq:PoisDisc}), and passing to a continuos-time description, where $\Delta t\to dt$ and $\Delta W^\rho_t\to dW^\rho_t$ become independent zero-mean Guassian white-noise increments, we arrive at the CLE in the It\=o sense~\cite{Gillespie2000,Gardiner}
\begin{equation}\label{eq:CLE}
\begin{split}
d{\bf X}_t  =\sum_{\rho\ge1} d{\bf X}^\rho_t=\sum_{\rho\ge 1} &\boldsymbol\nu^\rho (a_\rho({\bf X}_t)-a_{-\rho}({\bf X}_t))dt \\
 & + \boldsymbol\nu^\rho \sqrt{a_\rho({\bf X}_t)+a_{-\rho}({\bf X}_t)}dW^\rho_t.
\end{split}
 \end{equation}
The corresponding Stratonovich version of the CLE will also be useful
\begin{equation}\label{eq:CLEstrat}
\begin{split}
d{\bf X}_t=\sum_{\rho\ge1} d{\bf X}^\rho_t= \sum_{\rho\ge 1}\boldsymbol\nu^\rho \Big(&a_\rho({\bf X}_t)-a_{-\rho}({\bf X}_t) \\
&- \bnu^{\rho}\cdot\partial_{\bf x}(a_\rho({\bf X}_t)+a_{-\rho}({\bf X}_t))/2\Big)dt  
\\ &+ \boldsymbol\nu^\rho \sqrt{a_\rho({\bf X}_t)+a_{-\rho}({\bf X}_t)}\circ dW^\rho_t,
\end{split}
\end{equation}
where ``$\circ$'' denotes a Stratonovich intergral.
 Notice that we have combined the noises of the forward and reverse reactions of each channel into one Gaussian increment, so that $d{\bf X}_t^\rho$ represents the contribution to the net infinitesimal change in the molecular populations due to just reaction channel $\rho$.
 As we will see, this is the proper level of description that will allow us to connect the entropy flow of the CME to the CLE.

 An enlightening formulation of the CLE that will prove useful is to restructure it in a manner akin to the CME in Eq.~(\ref{eq:Poisson}).
Observe that Eq.~(\ref{eq:GausApprox}) allows us to identify 
 \begin{equation}\label{eq:GausApproxCLE}
 d{\mathcal N}^\rho_t = a_{\rho}({\bf X}_t)d t + \sqrt{a_\rho({\bf X}_t)}d W^\rho_t,
 \end{equation}
 as the number of  $R_\rho$ reactions in an infinitesimal interval $dt$ within the CLE limit.
 Thus, the flux -- the net number of reactions -- through reaction channel $\rho$ can be identified as $d{\mathcal N}^\rho_t-d{\mathcal N}^{-\rho}_t$, and the CLE can alternatively be  structured as [cf.~Eq.~(\ref{eq:Poisson})]
 \begin{equation}\label{eq:CLEjump}
 \begin{split}
 d{\bf X}_t &= \sum_{\rho\ge 1}d{\bf X}^\rho_t=\sum_{\rho\ge 1}\boldsymbol\nu^\rho(d{\mathcal N}^\rho_t-d{\mathcal N}^{-\rho}_t).
\end{split}
 \end{equation} 

With the CLE in hand, we can readily apply the formulation of stochastic thermodynamics for diffusion processes~\cite{Seifert2005b,VandenBroeck2010}, which first requires correctly identifying each constitutive microscopic process, that is the physical mechanisms that mediate changes in the populations.
For example, we could treat each reaction as a separate mechanism, or we could treat all the reactions as one mechanism, only tracking the total changes in the populations.
For a chemical reaction network, the correct level of description is to count each reaction channel separately, which we have anticipated with our formulation of the CLE in Eq.~(\ref{eq:CLE}).
With this in mind, the dynamic contribution to the entropy flow is identified as the ratio of the force to the diffusion coefficient in the Stratonovich version of the CLE [Eq.~(\ref{eq:CLEstrat})],
\begin{equation}\label{eq:CLEent}
\begin{split}
d\sigma^{\rm CLE}_t &= \sum_{\rho\ge 1} \sigma^{\rm CLE}_\rho({\bf X}_t)\circ\frac{1}{\boldsymbol\nu^\rho}\cdot d{\bf X}^\rho_t  \\ 
&= \sum_{\rho\ge 1} \sigma^{\rm CLE}_\rho({\bf X}_t)\circ (d{\mathcal N}^\rho_t-d{\mathcal N}^{-\rho}_t),
\end{split}
\end{equation}
with
\begin{equation}
\sigma^{\rm CLE}_\rho({\bf x})=\frac{a_\rho({\bf x})-a_{-\rho}({\bf x})-\boldsymbol\nu^\rho\cdot\partial_{\bf x}(a_\rho({\bf x})+a_{-\rho}({\bf x}))/2}{(a_\rho({\bf x})+a_{-\rho}({\bf x}))/2}.
\end{equation} 
Strikingly, the entropy flow cannot be connected to the heat: in the large population limit (${\bf x}\gg\bnu^\rho$), the local detailed balance relation [Eq.~(\ref{eq:LDB})] would read $\ln(a_\rho({\bf x})/a_{-\rho}({\bf x}))=-\beta\Delta\Phi_\rho({\bf x})$, but as the entropy flow is not expressible as a simple ratio of forward to reverse propensities, local detailed balance fails to provide the required connection.
We will see though that near equilibrium $\sigma^{\rm CLE}_\rho$ does approximate $\sigma^{\rm CME}_\rho$.
Note also, Xiao \emph{et al.~}in their analysis of the CLE utilized a coarser version of the entropy flow that does not distinguish the various reaction channels~\cite{Xiao2009}, underestimating the true entropy flow.

 \subsection{Low Noise Approximation}\label{sec:LNAdyn}
 
 Van Kampen's system size expansion is based on the observation that when the system size $\Omega$ is large, the typical behavior of the system is Gaussian~\cite{VanKampen}.
 Formally, we split the populations as ${\bf X}_t =\Omega{\boldsymbol\psi}_t+ \Omega^{1/2}{\boldsymbol\xi}_t$ in terms of a macroscopic, deterministic concentration $\boldsymbol\psi_t$ and small fluctuations $\boldsymbol\xi_t$.
The LNA then provides the dynamics for $\boldsymbol\psi_t$ and $\boldsymbol\xi_t$.
 However, to make a comparison with the CME and CLE, we choose a representation in terms of the macroscopic populations $\boldsymbol\phi_t=\Omega\boldsymbol\psi_t$ and ${\bf X}_t$:
  \begin{equation}\label{eq:LNADet}
 \dot{\boldsymbol\phi_t}=\sum_{\rho\ge 1} \boldsymbol\nu^\rho (a_\rho(\boldsymbol\phi_t)-a_{-\rho}(\boldsymbol\phi_t)),
 \end{equation}
 which is the macroscopic reaction rate equations, and
 \begin{equation}\label{eq:LNA}
 \begin{split}
 d{\bf X}_t = \sum_{\rho\ge 1} &d{\bf X}^\rho_t \\
 = \sum_{\rho\ge 1}& \boldsymbol\nu^\rho (a_\rho(\boldsymbol\phi_t)-a_{-\rho}(\boldsymbol\phi_t)) dt\\
 &+\boldsymbol\nu^\rho({\bf X}_t-\boldsymbol\phi_t) \cdot\partial_{\boldsymbol\phi}  (a_\rho({\boldsymbol\phi}_t)-a_{-\rho}(\boldsymbol\phi_t))dt \\
 & + {\boldsymbol\nu^\rho}\sqrt{a_\rho({\boldsymbol\phi}_t)+a_{-\rho}({\boldsymbol\phi_t})}dW^\rho_t.
 \end{split}
 \end{equation}
 Notice, we have expressed the dynamics as a sum of reaction channels in anticipation of the stochastic thermodynamics.
 
 We also observe that, like the CLE, the changes in ${\bf X}_t$ occur in discrete amounts given by $\boldsymbol\nu^\rho$ and are separately due to each reaction.
 Thus, we can identify the approximate number of $R_\rho$ reactions in the LNA as [cf.~Eq.~(\ref{eq:GausApproxCLE})]
 \begin{equation}
 \begin{split}
 d{\mathcal M}^\rho_t = \big[a_\rho(\boldsymbol\phi_t)+({\bf X}_t-\boldsymbol\phi_t) \cdot &\partial_{\boldsymbol\phi}  a_\rho({\boldsymbol\phi}_t)\big]dt \\
 & + \sqrt{a_\rho({\boldsymbol\phi}_t))}dW^\rho_t.
\end{split}
 \end{equation}
 Consequently, we can rewrite the LNA in Eq.~(\ref{eq:LNA}) as
 \begin{equation}\label{eq:LNAjump}
 d{\bf X}_t = \sum_{\rho\ge 1}d{\bf X}^\rho_t = \sum_{\rho\ge 1} \boldsymbol\nu^\rho(d{\mathcal M}^\rho_t -d{\mathcal M}^{-\rho}_t).
 \end{equation}

 Again, we apply the stochastic thermodynamics of diffusion processes to obtain the dynamic entropy flow as~\cite{Seifert2005b,VandenBroeck2010}
  \begin{equation}\label{eq:LNAent}
\begin{split}
d\sigma^{\rm LNA}_t&=\sum_{\rho\ge 1} \sigma^{\rm LNA}_\rho({\bf X}_t,\boldsymbol\phi_t)\circ \frac{1}{\boldsymbol\nu^\rho}\cdot d{\bf X}^\rho_t \\
&=\sum_{\rho\ge 1} \sigma^{\rm LNA}_\rho({\bf X}_t,\boldsymbol\phi_t)\circ (d{\mathcal M}^\rho_t -d{\mathcal M}^{-\rho}_t) 
\end{split}
\end{equation}
with
\begin{equation}\label{eq:LNAdynEnt}
\begin{split}
&\sigma^{\rm LNA}_\rho({\bf x},\boldsymbol\phi)\\
&=\sum_{\rho\ge 1}\frac{ a_\rho(\boldsymbol\phi)-a_{-\rho}(\boldsymbol\phi)+({\bf x}-\boldsymbol\phi)\cdot\partial_{\boldsymbol\phi} (a_\rho({\boldsymbol\phi})-a_{-\rho}(\boldsymbol\phi))}{(a_\rho(\boldsymbol\phi)+a_{-\rho}(\boldsymbol\phi))/2}.
\end{split}
\end{equation}
This entropy flow also cannot be expressed as a simple ratio of propensities, and therefore cannot be connected to the heat using local detailed balance [Eq.~(\ref{eq:LDB})].
 
 \subsection{Reaction rate equations}
 
 For exteremly large populations,  the fluctuations  about the mean (deterministic) solution become negligible.
 In this limit, the dynamics are governed by the deterministic macroscopic reaction rate equations (RRE)~\cite{VanKampen,Gillespie2000} 
  \begin{equation}\label{eq:RRE}
 \dot{\bf X}_t = \sum_{\rho\ge 1}\boldsymbol\nu^\rho(a_\rho({\bf X}_t)-a_{-\rho}({\bf X}_t)),
 \end{equation}
obtained as the limiting equation of either the CLE or LNA by dropping the noise.
The steady-state populations of the RRE, ${\bf X}^{\rm ss}$, are the solution of
\begin{equation}
\sum_{\rho\ge 1}\boldsymbol\nu^\rho(a_\rho({\bf X}^{\rm ss})-a_{-\rho}({\bf X}^{\rm ss}))=0.
\end{equation}
Like the CME, when each reaction is counter-balanced by its reverse, 
\begin{equation}\label{eq:RREdb}
a_{\rho}({\bf X}^{\rm ss}) = a_{-\rho}({\bf X}^{\rm ss}),
\end{equation}
we will say that the RRE is detailed balanced and identify the steady-state as equilibrium, ${\bf X}^{\rm ss} = {\bf X}^{\rm eq}$.
Crucially, the definition of detailed balance for the CME in Eq.~(\ref{eq:DB}) is equivalent to the definition for the RRE in the large population limit (${\bf X}\gg \boldsymbol\nu^\rho$).

 \section{Consistency at Equilibrium}\label{sec:equilibrium}
 
 The expressions of the entropy flows for the diffusion approximations differ dramatically from the CME's.
 In this section, we provide simple arguments that demonstrate that the three notions of entropy flow coincide only near equilibrium.
 These arguments are corroborated in the Appendices, where systematic expansions are applied to the generating function for entropy-flow fluctuations in the CME to show the CME's entropy-flow distribution collapses onto the diffusion approximations' entropy-flow distributions near equilibrium.
 
  \subsection{Chemical Langevin Equation}\label{sec:CLEapprox}
 
 We wish to demonstrate that the entropy flow in the CLE is approximately equal to the CME's, near equilibrium.
 To this end, we show that $  \sigma^{\rm CME}_\rho\approx\sigma^{\rm CLE}_\rho$ for each reaction channel, when ${\bf X}_t\approx{\bf X}^{\rm eq}$.
 
 In the CLE limit, the propensities are insensitive to small changes of size $\boldsymbol\nu^\rho$ in the populations.
Thus, we can expand $\sigma^{\rm CME}_\rho$ in Eq.~(\ref{eq:CMEdynEnt}) as
\begin{align}
\sigma^{\rm CME}_{\rho}({\bf X}_t) & = \ln \frac{a_\rho({\bf X}_t)}{a_{-\rho}({\bf X}_t+\boldsymbol\nu^\rho)} \\
&\approx \ln \frac{a_\rho({\bf X}_t)}{a_{-\rho}({\bf X}_t)}-\frac{\boldsymbol\nu^\rho\cdot\partial_{\bf x}a_{-\rho}({\bf X}_t)}{a_{-\rho}({\bf X}_t)}.
\end{align}
The first term is the most problematic as there are no logarithms in $\sigma^{\rm CLE}_\rho$.
However, this term is small near equilibrium.
Specifically for large, detailed-balanced systems, we expect the typical population to be near the steady-state equilibrium population ${\bf X}_t\approx{\bf X}^{\rm eq}$, characterized by $a_{\rho}({\bf X}^{\rm eq})=a_{-\rho}({\bf X}^{\rm eq})$.
Therefore, (see Appendix~\ref{sec:AppCLE})
\begin{equation}
\ln \frac{a_\rho({\bf X}_t)}{a_{-\rho}({\bf X}_t)}\approx\frac{a_{\rho}({\bf X}_t)-a_{-\rho}({\bf X}_t)}{a_{\rho}({\bf X}_t)}.
\end{equation}
 Finally, within the same level of approximation, we have $a_{\pm\rho}\approx (a_\rho+a_{-\rho})/2$, and
 \begin{equation}
 \begin{split}
\sigma^{\rm CME}_{\rho}({\bf X}_t)\approx&\frac{a_\rho({\bf X}_t)-a_{-\rho}({\bf X}_t)}{(a_\rho({\bf X}_t)+a_{-\rho}({\bf X}_t))/2} \\
&-\frac{\boldsymbol\nu^\rho\cdot\partial_{\bf x}(a_\rho({\bf X}_t)+a_{-\rho}({\bf X}_t))}{a_\rho({\bf X}_t)+a_{-\rho}({\bf X}_t)},
\end{split}
\end{equation} 
 which is equivalent to $\sigma^{\rm CLE}_\rho$.
 Consequently, we have the expected equivalence $ ds^{\rm e,CLE}_t\approx ds^{\rm e,CME}_t$,
 valid for a detailed-balanced system in the equilibrium steady state.
 This correspondence confirms that we have correctly identified the necessary constitutive processes for tracking the entropy production.
 
 \subsection{Low Noise Approximation}\label{sec:LNAapprox}
 
 Like the CLE, $\sigma^{\rm CME}_\rho$ can be approximated in the LNA  limit as $\sigma^{\rm LNA}_\rho$.
 In the LNA, we split the populations as ${\bf X}_t=\Omega\boldsymbol\psi_t+\Omega^{1/2}\boldsymbol\xi_t$.
 The approximation is then facilitated by expressing $\sigma_\rho^{\rm CME}$ in terms of the concentrations through the relation $a_\rho({\bf X}_t)=\Omega \alpha_\rho({\bf X}_t/\Omega)$ as 
 \begin{align}\nonumber
 \sigma^{\rm CME}_{\rho}&({\bf X}_t) \\
 &   = \ln \frac{\alpha_\rho(\boldsymbol\psi_t+\boldsymbol\xi_t/\Omega^{1/2})}{\alpha_{-\rho}(\boldsymbol\psi_t+\boldsymbol\xi_t/\Omega^{1/2}+\boldsymbol\nu^\rho/\Omega)} \\
 &\approx \ln\frac{\alpha_\rho(\boldsymbol\psi_t)}{\alpha_{-\rho}(\boldsymbol\psi_t)} +\frac{1}{\Omega^{1/2}}\bxi\cdot\partial_{\boldsymbol\psi}\ln\frac{\alpha_\rho(\boldsymbol\psi_t)}{\alpha_{-\rho}(\boldsymbol\psi_t)} \\
  &= \ln\frac{a_\rho(\boldsymbol\phi_t)}{a_{-\rho}(\boldsymbol\phi_t)} +({\bf X}_t-\bphi_t) \cdot\partial_{\boldsymbol\phi}\ln\frac{a_\rho(\boldsymbol\phi_t)}{a_{-\rho}(\boldsymbol\phi_t)}.
  \end{align}
 We obtain a connection with the LNA entropy flow when the dynamics are detailed balanced, and we are in the time-independent, equilibrium steady state, $\boldsymbol\phi_t=\boldsymbol\phi^{\rm eq}$ with $a_\rho(\bphi^{\rm eq})=a_{-\rho}(\bphi^{\rm eq})$.
  At equilibrium then
  \begin{equation}
   \sigma^{\rm CME}_{\rho}({\bf X}_t) \approx ({\bf X}_t-\bphi^{\rm eq}) \cdot\partial_{\boldsymbol\phi}\ln\frac{a_\rho(\boldsymbol\phi)}{a_{-\rho}(\boldsymbol\phi)}\bigg|_{\bphi^{\rm eq}} = \sigma^{\rm CLE}_\rho({\bf X}_t,\bphi^{\rm eq}),
  \end{equation}
  which can be verified by evaluating Eq.~(\ref{eq:LNAdynEnt}) at $\bphi_t=\bphi^{\rm eq}$.
  Thus, again near equilibrium,  $ds^{\rm e,LNA}_t\approx ds^{\rm e,CME}_t$.
 
\section{Approximating the entropy flow}\label{sec:approx}

We have seen that the expressions for the entropy flows of the CLE and LNA have no connection to the entropy flow of the CME, except at equilibrium.
However, the population dynamics of the CLE and LNA both approximate the dynamics of the CME for large populations~\cite{Gillespie2000,Gillespie2002}.
One may wonder then if we can use that correspondence to at least approximate  the CME's entropy-flow fluctuations using the diffusion approximations.
In this section, we provide such an approximation, and later verify it numerically with an example in Sec.~\ref{sec:example}.
As a consequence, the local detailed balance relation allows us to use the CLE and LNA to approximate the CME's heat fluctuations as well.  

\subsection{Chemical Langevin Equation}
 
To approximate the entropy flow in the CME using the CLE, let us begin by observing that in the large population limit (${\bf X}\gg \bnu^\rho$) away from equilibrium we can approximate the dynamic entropy flow of the CME [Eq.~(\ref{eq:CMEdynEnt})] as
\begin{equation}
\sigma^{\rm CME}_\rho({\bf X}_t) \approx - \sigma^{\rm CME}_{-\rho}({\bf X}_t)\approx \ln\frac{a_{\rho}({\bf X}_t)}{a_{-\rho}({\bf X}_t)} .
\end{equation}
Thus, in every reaction channel there is a fixed entropy flow.
Noting from Eq.~(\ref{eq:CLEjump}) that in the CLE the flux through that channel is $d{\mathcal N}^\rho_t-d{\mathcal N}^{-\rho}_t$, we have the approximate entropy flow,
\begin{equation}\label{eq:CLEentApprox}
d\sigma^{\rm CME}_t\approx \sum_{\rho\ge 1}\ln\frac{a_{\rho}({\bf X}_t)}{a_{-\rho}({\bf X}_t)} \left(d{\mathcal N}^\rho_t-d{\mathcal N}^{-\rho}_t\right),
\end{equation}
reminiscent of Eq.~(\ref{eq:CMEentDyn}).
Importantly, this expression can be evaluated using only the solution to the CLE, and is clearly connected to the heat flow through the local detailed balance relation.

\subsection{Low Noise Approximation}
 
 Within the LNA, the coarsest approximation to the CME entropy flow in each reaction is 
 \begin{equation}
\sigma^{\rm CME}_\rho({\bf X}_t) \approx -\sigma^{\rm CME}_{-\rho}({\bf X}_t) \approx \ln\frac{a_\rho(\bphi_t)}{a_{-\rho}(\bphi_t)},
 \end{equation}
 which is dominated by the deterministic macroscopic dynamics.
 Observing from Eq.~(\ref{eq:LNAjump}) that the flux through reaction channel $\rho$ within the LNA is $d{\mathcal M}^\rho_t-d{\mathcal M}^{-\rho}_t$, we have the LNA approximation to the CME entropy flow,
\begin{equation}\label{eq:LNAentApprox}
 d\sigma^{\rm CME}_t \approx\sum_{\rho\ge 1}\ln\frac{a_\rho(\bphi_t)}{a_{-\rho}(\bphi_t)}\left(d{\mathcal M}^\rho_t - d{\mathcal M}^{-\rho}_t\right),
\end{equation}
just like Eq.~(\ref{eq:CMEentDyn}).
Crucially this approximation can be evaluated solely with knowledge of the solution to the LNA.

\subsection{Reaction Rate Equations}

In the extreme large population limit, the noise becomes negligible and the dynamics follow the deterministic RRE.
When the RRE is valid, the mean behavior becomes the typical behavior.
Thus, within the RRE approximation, $d{\mathcal N}^\rho_t\approx d{\mathcal M}^\rho_t\approx a_{\rho}({\bf X}_t)dt$, and we can readily approximate the CME entropy flow as
\begin{equation}\label{eq:RREentApprox}
d\sigma^{\rm CME}_t \approx  \sum_{\rho\ge 1}\ln\frac{a_{\rho}({\bf X}_t)}{a_{-\rho}({\bf X}_t)} \left(a_{\rho}({\bf X}_t)-a_{-\rho}({\bf X}_t)\right)dt,
\end{equation}
along the solution of the RRE in Eq.~(\ref{eq:RRE}).
Remarkably, we recover an expression for the entropy flow that is exactly the one proposed to study the irreversible thermodynamics of deterministic, chemical reaction networks~\cite{Polettini2014}.

 \section{Illustrative Example}\label{sec:example}
 Consider a single molecular species $A$ coupled to two particle reservoirs (or chemostats) with fixed chemical potentials $\mu_1$ and $\mu_2$. 
 An instructive way to formulate this setup is to recognize that each reservoir is attempting to impose its own equilibrium population $\omega_i=\exp(\beta\mu_i)$, for $i=1,2$.
Particles are then created and destroyed by exchange with the two reservoirs through the two pairs of reactions 
 \begin{equation}
 \begin{array}{l}
 \emptyset \xrightarrow{k\omega_1} A \\
  A\xrightarrow{kA} \emptyset \end{array}
\qquad
\begin{array}{l}  \emptyset \xrightarrow{k\omega_2} A \\ A\xrightarrow{kA} \emptyset  \end{array},
 \end{equation}
 with rate constant $k$.
 Steady-state is obtained  with mean population ${\bar A} = (\omega_1+\omega_2)/2$.
 
 For each of the three equations of motion -- the CME 
 \begin{equation}
 dA_t=\sum_{i=1}^2 k\omega_idN^i_t-kA_tdN^{-i}_t,
 \end{equation}
 the CLE,
 \begin{equation}
 dA_t=\sum_{i=1}^2 (k\omega_i-kA_t)dt+\sqrt{k\omega_i+kA_t}dW^i_t,
 \end{equation}
 and the LNA,
 \begin{equation}
  dA_t=\sum_{i=1}^2 (k\omega_i-kA_t)dt+\sqrt{k\omega_i+k{\bar A}}dW^i_t
 \end{equation}
 -- we have simulated 1000 steady-state trajectories of length $\tau = 50 s$ with $k=1s^{-1}$  under nonequilibrium conditions with $\omega_1=500$ and $\omega_2=100$ and equilibrium conditions with $\omega_1=\omega_2=100$ (as described in the Methods).
In both scenarios, the steady state is well within the large population limit with ${\bar A}\gg 1$.
For each trajectory, we calculated the time-averaged dynamic entropy flow $\dot\sigma = \frac{1}{\tau}\int_0^\tau d\sigma_s$, using the appropriate formula:
Eq.~(\ref{eq:CMEentDyn}) for the CME with 
 \begin{equation}
 \sigma_i^{\rm CME}(A)= -\sigma^{\rm CME}_{-i}(A+1)= \ln\left(\frac{\omega_i}{A+1}\right),
 \end{equation}
  $i=1,2$;
Eq.~(\ref{eq:CLEent}) for the CLE with 
 \begin{equation}
\sigma_i^{\rm CLE}(A) = \frac{2(\omega_i-A-1/2)}{\omega_i+A};
 \end{equation}
 and Eq.~(\ref{eq:LNAent}) for the LNA with
 \begin{equation}
 \sigma^{\rm LNA}_i= \frac{2(\omega_i-A)}{\omega_i+{\bar A}}.
 \end{equation}
From those values we constructed histograms for the probability distributions of entropy flows.

\begin{figure}[htb]
\centering
\includegraphics[scale=.35]{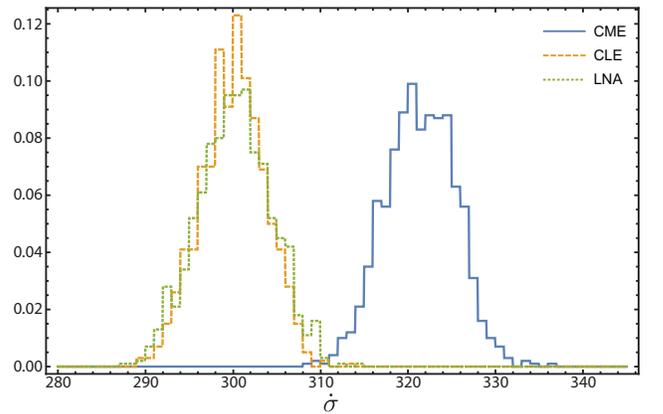}
\caption{Probability distribution of the entropy flow rate in a nonequilibrium steady state for the CME (blue solid), CLE (orange dashed), and LNA (green dotted).}
\label{fig:NoneqEnt}
\end{figure}

 In Fig.~(\ref{fig:NoneqEnt}), we have plotted the entropy-flow distribution under nonequilibrium conditions.
The entropy flows for the diffusion approximations are similar to each other, but significantly different from the CME entropy flow.
 Specifically, the mean values of the entropy flow of the CLE and LNA lower bound the mean in the CME.
 This is to be expected, as the CLE and LNA, are coarse-grained descriptions of the dynamics, and coarse graining very generally decreases the mean entropy production~\cite{Kawai2007,Parrondo2009,Gomez-Marin2008b,Horowitz2009b,Esposito2012c}.

\begin{figure}[htb]
 \centering
 \includegraphics[scale=.35]{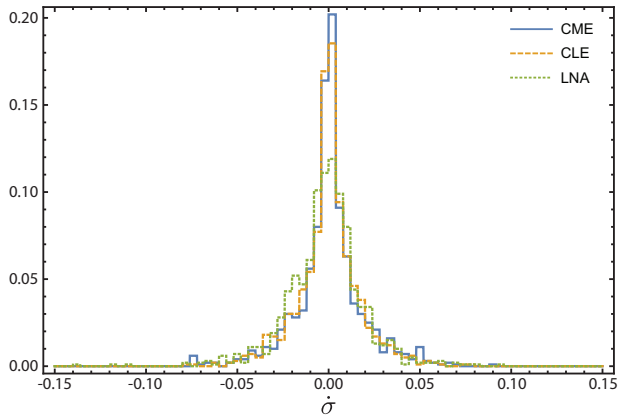}
 \caption{Probability distribution of the entropy flow rate in an equilibrium steady state for the CME (blue solid), CLE (orange dashed), and LNA (green dotted).}
 \label{fig:EqEnt}
 \end{figure}

By contrast, the equilibrium entropy flows of the three descriptions collapse onto each other, as demonstrated in Fig.~(\ref{fig:EqEnt}).
 While the entropy flow of the CLE seems to approximate well the CME's, the LNA has slightly wider tails.
 This is not too surprising: the CLE and LNA are Gaussian approximations, so they get the mean and variance of the populations correct, but poorly estimate the tails as confirmed by direct simulation~\cite{Gillespie2002}.

\begin{figure}[htb]
\includegraphics[scale=.35]{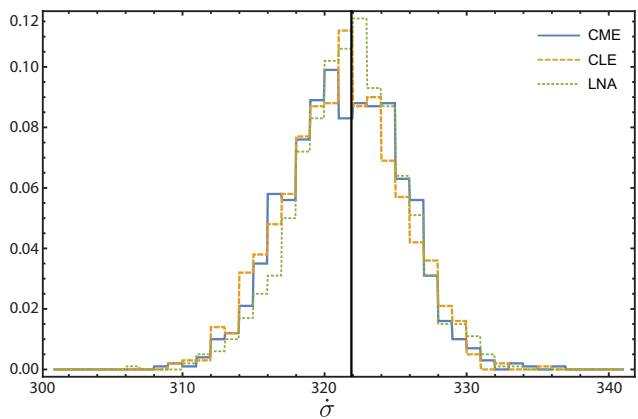}
\caption{Diffusion approximations to the probability distribution of entropy flow in the CME (solid blue) away from equilibrium using the CLE (dashed orange) and LNA (dotted green). The black vertical line denotes the macroscopic entropy production obtained from the RRE, $\dot\sigma^{\rm RRE}\approx 322$.}
\label{eq:DiffApprox}
\end{figure}

Finally, using Eqs.~(\ref{eq:CLEentApprox}) and (\ref{eq:LNAentApprox}), we have constructed histograms of the underlying CME entropy-flow distribution using the diffusion approximations.
The results appear in Fig.~(\ref{eq:DiffApprox}).
The diffusion approximations offer remarkably good approximations of the CME entropy-flow fluctuations.
For comparison, we have included the deterministic entropy flow obtained from the RRE in Eq.~(\ref{eq:LNAentApprox}) as the vertical black line at $\dot\sigma^{\rm RRE}\approx 322$, which as expected falls at the mean. 
 
  \section{Conclusion}\label{sec:conclusion}
  
  In general, the CLE and LNA do not have a consistent stochastic thermodynamics, as their dynamic entropy flow is unrelated to the energetics inherited from the CME.
  Near equilibrium, however, all the different notions of heat and entropy flow collapse, and we recover a consistent stochastic thermodynamics for the diffusion approximations.
  This property stems from the fact that the diffusion approximations arise as a coarse graining in time of the CME~\cite{Parrondo2009,Gomez-Marin2008b,Horowitz2009b}, where we smear out the dynamics onto a longer time scale.
  Generically, when we coarse grain we throw away information about precisely which trajectories the system follows, and with that we loose the ability to accurately determine the entropy flow~\cite{Kawai2007,Santillan2011,Esposito2012c,Sekimoto,Roldan2012}.
  However, near equilibrium each chemical reaction contributes little to the entropy flow due to detailed balance.
  As such, by coarse graining over the precise sequence of chemical reactions, we do not loose any substantial information about the entropy production.
  As a result, the CLE and LNA entropy flow are very close to the underlying entropy flow of the CME.
  Generically, we suspect that these conclusions about entropy flows remain valid, whenever we coarse grain in time.

  Away from equilibrium, we found that we can still use the diffusion approximations to estimate the entropy-flow fluctuations in the CME.
  This observation implies that we can access the true thermodynamic character of the underlying chemical reaction network using the diffusion approximations without having to solve the CME, as long as we are in the appropriate limit.
  Such an approach offers a great simplification in the analysis of the thermodynamics and energetics, in addition to the dynamics.

\section*{Methods}

Sample trajectories for the CME were generated using Gillespie's algorithm~\cite{Breuer}, for the CLE using a stochastic Runge-Kutta algorithm~\cite{SanMiguel}, and for the LNA using the stochastic Heun method~\cite{Sekimoto}.
  
 \begin{acknowledgments}
I am very grateful to Matteo Polettini and Massimiliano Esposito for insightful discussions and suggestions.  I would also like to acknowledge Jason Green for reviewing this manuscript.
This work was financially supported by the Spanish Government, grant ENFASIS (FIS2011-22644), by the National Research Fund of Luxembourg (project FNR/A11/02), and by the NSF project PHY-1212413.
\end{acknowledgments}

 \begin{widetext}
 \appendix

 \section{Entropy flow fluctuations in the CLE limit} \label{sec:AppCLE}
 
 In this appendix, we show that the entropy-flow fluctuations of the CME only agree with those of the CLE near chemical equilibrium.  
 To access the entropy-flow fluctuations, we follow the method developed by Imparato and Peliti~\cite{Imparato2005c} and begin by considering the Fokker-Planck equation for the populations and total dynamic entropy flow $\sigma=\int_0^t d\sigma_s^{\rm CME}$ up to time $t$:
 \begin{equation}\label{eq:FPxent}
  \frac{\partial }{\partial t}P_t({\bf x},\sigma) =\sum_{\rho=\pm 1}^{\pm M}   a_\rho({\bf x}-{\boldsymbol\nu}^\rho)P_t\left({\bf x}-\boldsymbol\nu^\rho,\sigma-\sigma^{\rm CME}_{\rho}({\bf x}-\boldsymbol\nu^\rho)\right)
-a_\rho({\bf x}) P_t({\bf x},\sigma),
 \end{equation}
 The analysis of this equation is facilitated by switching to the generating function for the entropy flow $G_t({\bf x},\lambda)=\langle e^{-\lambda \sigma}\rangle_t$, whose equation of motion is obtained from Eq.~(\ref{eq:FPxent}), 
% \begin{equation}
% \begin{split}
%  \frac{\partial }{\partial t}G_t({\bf x},\lambda) =\sum_{\rho}   a_\rho({\bf x}-{\boldsymbol\nu}^\rho)e^{-\lambda q_\rho({\bf x}-\boldsymbol\nu^\rho)}&G_t({\bf x}-\boldsymbol\nu^\rho,\lambda) \\
%  &-a_\rho({\bf x}) G_t({\bf x},\lambda).
%  \end{split}
% \end{equation}
  \begin{equation}\label{eq:CLEgenFunc}
  \frac{\partial }{\partial t}G_t({\bf x},\lambda) =\sum_{\rho=\pm 1}^{\pm M}  [a_{-\rho}({\bf x})]^{\lambda} [a_\rho({\bf x}-{\boldsymbol\nu}^\rho)]^{-\lambda+1}G_t({\bf x}-\boldsymbol\nu^\rho,\lambda) -a_\rho({\bf x}) G_t({\bf x},\lambda),
 \end{equation}
 where we have used the definition $\sigma^{\rm CME}_\rho = \ln a_\rho/a_{-\rho}$ [Eq.~(\ref{eq:CMEdynEnt})].
 Now as Gillespie has observed, we can obtain the CLE by using the Kramers-Moyal expansion of the Fokker-Planck equation~\cite{Gillespie2000,Risken}.
 To track this expansion, let us explicitly introduce a small parameter through the substitution $\boldsymbol\nu^\rho\to\epsilon\boldsymbol\nu^\rho$, which formalizes the idea that the change in populations during the reactions are small.
 We can set $\epsilon=1$ at the end of the calculation.
 Expanding Eq.~(\ref{eq:CLEgenFunc}) to second order in $\epsilon$ and changing the sum over reactions to a sum over reaction channels, we obtain
%  \begin{equation}
% \begin{split}
%  \frac{\partial }{\partial t}G_t({\bf x},\lambda) =\sum_{\rho=\pm 1}^{\pm M} &a_{\rho}({\bf x})\left[\left(\frac{a_{-\rho}({\bf x})}{a_{\rho}({\bf x})}\right)^\lambda-1\right]G_t({\bf x},\lambda) \\
%  &-\epsilon [a_{-\rho}({\bf x})]^\lambda(\bnu^\rho\cdot\partial_{\bf x})[a_\rho({\bf x})]^{-\lambda+1}G_t({\bf x},\lambda) +\frac{\epsilon^2}{2}[a_{-\rho}({\bf x})]^\lambda(\boldsymbol\nu^\rho\cdot\partial_{\bf x})^2[a_\rho({\bf x})]^{-\lambda+1} G_t({\bf x},\lambda).
%  \end{split}
% \end{equation} 
% Anticipating the final result, we change the sum over reactions to the sum over reactions channels
   \begin{equation}
 \begin{split}
  \frac{\partial }{\partial t}G_t({\bf x},\lambda) =\sum_{\rho\ge 1}  &a_{\rho}({\bf x})\left[\left(\frac{a_{-\rho}({\bf x})}{a_{\rho}({\bf x})}\right)^\lambda-1\right]G_t({\bf x},\lambda) + a_{-\rho}({\bf x})\left[\left(\frac{a_{\rho}({\bf x})}{a_{-\rho}({\bf x})}\right)^\lambda-1\right]G_t({\bf x},\lambda) \\
  &-\epsilon\bnu^\rho\cdot \Big\{[a_{-\rho}({\bf x})]^\lambda\partial_{\bf x}[a_\rho({\bf x})]^{-\lambda+1}G_t({\bf x},\lambda)-[a_{\rho}({\bf x})]^\lambda\partial_{\bf x}[a_{-\rho}({\bf x})]^{-\lambda+1}G_t({\bf x},\lambda)\Big\} \\
  &+\frac{\epsilon^2}{2}[a_{-\rho}({\bf x})]^\lambda(\boldsymbol\nu^\rho\cdot\partial_{\bf x})^2[a_\rho({\bf x})]^{-\lambda+1} G_t({\bf x},\lambda).
  \end{split}
 \end{equation} 
 Next to formalize the near equilibrium approximation, we assume that each reaction is approximately balanced by its reverse
 \begin{equation}\label{eq:delta}
 a_{-\rho}({\bf x})=a_{\rho}({\bf x})\left(1+\epsilon\delta({\bf x})\right).
 \end{equation}
for some function $\delta$.
Expanding, we find after significant rearrangement
  \begin{equation}
 \begin{split}
\frac{\partial }{\partial t}G_t({\bf x},\lambda)& \\
  =\epsilon^2\sum_{\rho\ge 1}  & (\bnu^\rho\cdot\partial_{\bf x}) a_{\rho}({\bf x}) (\bnu^\rho\cdot\partial_{\bf x})G_t({\bf x},\lambda) \\
  & -\lambda a_{\rho}({\bf x})\left(\delta({\bf x})+\frac{\bnu^\rho\cdot\partial_{\bf x}a_{\rho}({\bf x})}{a_{\rho}({\bf x})}\right)(\bnu^\rho\cdot\partial_{\bf x})G({\bf x},\lambda) - (\lambda-1)(\bnu^\rho\cdot\partial_{\bf x})\left[a_{\rho}({\bf x})\left(\delta({\bf x})+\frac{\bnu^\rho\cdot\partial_{\bf x}a_{\rho}({\bf x})}{a_{\rho}({\bf x})}\right)G({\bf x},\lambda)\right]\\
  & +\lambda(\lambda-1) a_{\rho}({\bf x})\left(\delta({\bf x})+\frac{\bnu^\rho\cdot\partial_{\bf x}a_{\rho}({\bf x})}{a_{\rho}({\bf x})}\right)^2 G({\bf x},\lambda)\\
  & \equiv {\mathcal L}(\lambda)G({\bf x},\lambda).
  \end{split}
 \end{equation} 
 This equation describes the joint dynamics of a Stratonovich diffusion process and its entropy flow~\cite{Imparato2007}.
 One way to verify this fact is to note that the generator of the dynamics ${\mathcal L}$ satisfies the symmetry ${\mathcal L}(1-\lambda)={\mathcal L}^\dag(\lambda)$, which guarantees the entropy flow verifies a fluctuation theorem~\cite{Imparato2007}.
 The corresponding Langevin equations are
 \begin{align}
 d{\bf X}_t &= \sum_{\rho\ge 1} - \bnu^\rho a_\rho({\bf X}_t)\left(\delta({\bf X}_t)+\frac{\bnu^\rho\cdot\partial_{\bf x}a_{\rho}({\bf X}_t)}{a_{\rho}({\bf X}_t)}\right)dt +\bnu^\rho \sqrt{2a_{\rho}({\bf X}_t)}\circ dW^{\rho}_t \\
 d\sigma_t&=\sum_{\rho\ge 1}-\left(\delta({\bf X}_t)+\frac{\bnu^\rho\cdot\partial_{\bf x}a_{\rho}({\bf X}_t)}{a_{\rho}({\bf X}_t)}\right)\circ \frac{1}{\bnu^\rho}\cdot d{\bf X}_t^\rho.
 \end{align}

 We can verify that this entropy flow provides the correct correspondence between the CME and CLE discussed in Sec.~\ref{sec:CLEapprox} by approximating the CME entropy flow to lowest order in $\epsilon$:
 \begin{align}
  \sigma^{\rm CME}_\rho ({\bf x}) &= \ln\frac{a_{\rho}({\bf x})}{a_{-\rho}({\bf x}+\bnu^\rho)} \\
  &\approx -\epsilon\left(\delta({\bf x})+\frac{\bnu^\rho\cdot\partial_{\bf x}a_{\rho}({\bf x})}{a_{\rho}({\bf x})}\right) \\
 & \approx \frac{a_{\rho}({\bf x})-a_{-\rho}({\bf x})}{a_{\rho}({\bf x})}-\epsilon\frac{\bnu^\rho\cdot\partial_{\bf x}a_{\rho}({\bf x})}{a_{\rho}({\bf x})} \\
 &\approx \frac{a_{\rho}({\bf x})-a_{-\rho}({\bf x})}{(a_{\rho}({\bf x})+a_{\rho}({\bf x}))/2}-\epsilon\frac{\bnu^\rho\cdot\partial_{\bf x}(a_{\rho}({\bf x})+a_{-\rho}({\bf x}))}{a_{\rho}({\bf x})+a_{-\rho}({\bf x})} = \sigma^{\rm CLE}_\rho({\bf x}),
 \end{align}
  using the near equality of propensities in Eq.~(\ref{eq:delta}).
  
  Thus, when we expand the joint population and entropy-flow dynamics of the CME using the CLE approximation and demand that the resulting dynamics correctly identifies the coarse-grained entropy-flow by enforcing the fluctuation theorem, we are forced to consider only near equilibrium dynamics.

  \section{Entropy flow fluctuations in the LNA limit}\label{sec:AppLNA}
  
  To address the entropy-flow fluctuations in the LNA, we begin as in Appendix \ref{sec:AppCLE} with the Fokker-Planck equation for populations and entropy flow, except with the system-size dependence explicit:
   \begin{equation}
 \begin{split}
  \frac{\partial }{\partial t}P_t({\bf x},\sigma) =\sum_{\rho=\pm 1}^{\pm M}   \Omega\alpha_\rho\left(\frac{{\bf x}-{\boldsymbol\nu}^\rho}{\Omega}\right)P_t({\bf x}-\boldsymbol\nu^\rho,\sigma-\sigma^{\rm CME}_{\rho}({\bf x}-\boldsymbol\nu^\rho)) -\Omega\alpha_\rho\left(\frac{{\bf x}}{\Omega}\right) P_t({\bf x},\sigma).
  \end{split}
 \end{equation}
 The corresponding entropy-flow generating function satisfies
   \begin{equation}
 \begin{split}
  \frac{\partial }{\partial t}G_t({\bf x},\lambda) =\sum_{\rho=\pm 1}^{\pm M}   \Omega\left[\alpha_{-\rho}\left(\frac{{\bf x}}{\Omega}\right)\right]^{\lambda} \left[\alpha_\rho\left(\frac{{\bf x}-{\boldsymbol\nu}^\rho}{\Omega}\right)\right]^{-\lambda+1}G_t({\bf x}-\boldsymbol\nu^\rho,\lambda) -\Omega \alpha_\rho\left(\frac{{\bf x}}{\Omega}\right) G_t({\bf x},\lambda).
  \end{split}
 \end{equation}
 At this point, we carry through the steps in developing the system-size expansion~\cite{VanKampen}.
 We split the populations as  ${\bf x} = \Omega \boldsymbol\psi_t+\Omega^{1/2}\boldsymbol\xi$ for some as yet unspecified function of time $\bpsi_t$, and introduce $G_t({\bf x},\lambda)=G_t(\Omega \boldsymbol\psi_t+\Omega^{1/2}\boldsymbol\xi,\lambda)\equiv\Gamma_t({\xi},\lambda)$.
 The differential equation for $\Gamma$ can then be found following the steps outlined in Ref.~\cite{VanKampen},
 \begin{equation}\label{eq:Gamma}
 \begin{split}
\dot\Gamma_t&({\xi},\lambda)-\Omega^{1/2}\dot{\boldsymbol\psi_t}\cdot\partial_{\boldsymbol\xi}\Gamma({\boldsymbol\xi},\lambda)  \\
 &=\sum_{\rho=\pm 1}^{\pm M} \Omega[\alpha_{-\rho}(\bpsi_t+\Omega^{-1/2}\bxi)]^\lambda[\alpha_\rho(\bpsi_t+\Omega^{-1/2}\bxi-\Omega^{-1}\bnu^\rho)]^{-\lambda+1}\Gamma(\bxi-\Omega^{-1/2}\bnu^\rho,\lambda) - \Omega\alpha_\rho(\bpsi_t+\Omega^{-1/2}\bxi)\Gamma(\bxi,\lambda)
 \end{split}
 \end{equation}
 We then expand for $\Omega\gg 1$ and compare terms order by order.
 
 \emph{Order $O(\Omega)$}: The highest order in $\Omega$ terms, after some simplification, are \begin{equation}
 \Gamma(\bxi,\lambda)\sum_{\rho=\pm 1}^{\pm M} \alpha_\rho(\bpsi_t)\left[\left(\frac{\alpha_{-\rho}(\bpsi_t)}{a_\rho(\bpsi_t)}\right)^\lambda-1\right]=0.
 \end{equation}
 This needs to be true for all $\lambda$, which demands that
 \begin{equation}\label{eq:LNA0}
 \alpha_{\rho}(\bpsi_t)=\alpha_{-\rho}(\bpsi_t),
 \end{equation}
 which is the detailed balance condition for the RRE [Eq.~(\ref{eq:RREdb})].
 
\emph{Order $O(\Omega^{1/2})$}: The next order terms can be simplified to
 \begin{equation}
 \left[\dot\bpsi_t-\sum_{\rho=\pm 1}^{\pm M}\bnu^\rho\alpha_\rho(\bpsi_t)\right]\cdot\partial_\bxi\Gamma(\bxi,\lambda) = \lambda\sum_{\rho=\pm 1}^{\pm M} \alpha_\rho(\bpsi_t)\left(\bxi\cdot\partial_\bpsi\ln\frac{\alpha_{-\rho}(\bpsi_t)}{\alpha_{\rho}(\bpsi_t)}\right)\Gamma(\bxi,\lambda)=0,
 \end{equation}
 using the equality in Eq.~(\ref{eq:LNA0}).
 Here, the sum proportional to $\lambda$ is zero, since $\ln a_\rho/a_{-\rho}$ is antisymmetric under the interchange $\rho\to-\rho$. 
 As a result, we conclude that the macroscopic contribution evolves according to the RRE,
 \begin{equation}
 \dot{\bpsi_t}=\sum_{\rho=\pm 1}^{\pm M}\bnu^\rho\alpha_\rho(\bpsi_t).
 \end{equation}
 This fact combined with the detailed balance requirement in Eq.~(\ref{eq:LNA0}) forces us to only use systems at equilibrium,  $\bpsi_t=\bpsi^{\rm eq}$.
 
 \emph{Order $O(1)$}: At this order, we have the LNA, which, after substantial simplification using $\bpsi_t=\bpsi^{\rm eq}$, we arrive at
 \begin{equation}
 \begin{split}
 \dot\Gamma(\bxi,\lambda) = \sum_{\rho\ge1}&\alpha_\rho(\bpsi^{\rm eq})(\bnu^\rho\cdot\partial_\bxi)^2\Gamma(\bxi,\lambda)  \\
 &+ (\lambda-1)\alpha_\rho(\bpsi^{\rm eq})(\bnu^\rho\cdot\partial_\bxi)\left[\left(\bxi\cdot\partial_\bpsi\ln\frac{\alpha_{\rho}(\bpsi^{\rm eq})}{\alpha_{-\rho}(\bpsi^{\rm eq})}\right)\Gamma(\bxi,\lambda)\right] 
 +\lambda\alpha_\rho(\bpsi^{\rm eq})\left(\bxi\cdot\partial_\bpsi\ln\frac{\alpha_{\rho}(\bpsi^{\rm eq})}{\alpha_{-\rho}(\bpsi^{\rm eq})}\right)(\bnu^\rho\cdot\partial_\bxi\Gamma(\bxi,\lambda))\\
& +\lambda(\lambda-1)\alpha_\rho(\bpsi^{\rm eq})\left(\bxi\cdot\partial_\bpsi\ln\frac{\alpha_{-\rho}(\bpsi^{\rm eq})}{\alpha_{\rho}(\bpsi^{\rm eq})}\right)^2\Gamma(\bxi,\lambda)\\
&\equiv{\mathcal L}(\lambda)\Gamma(\bxi,\lambda)
 \end{split}
 \end{equation}
 This equation represents the joint population entropy-flow dynamics for a diffusion process~\cite{Imparato2007};
notice that it has the symmetry ${\mathcal L}(1-\lambda)={\mathcal L}^\dag(\lambda)$, which is required for the entropy production to satisfy a fluctuation theorem~\cite{Imparato2007}. 
 The corresponding Langevin equations are
 \begin{align}
 d\bxi_t &= \sum_{\rho\ge 1}\bnu^\rho\left(\bxi\cdot\partial_\bpsi\ln\frac{\alpha_{\rho}(\bpsi^{\rm eq})}{\alpha_{-\rho}(\bpsi^{\rm eq})}\right)dt + \bnu^\rho\sqrt{2\alpha_\rho(\bpsi^{\rm eq})}dW^\rho_t \\
 d\sigma_t &= \sum_{\rho\ge 1} \left(\bxi\cdot\partial_\bpsi\ln\frac{\alpha_{\rho}(\bpsi^{\rm eq})}{\alpha_{-\rho}(\bpsi^{\rm eq})}\right)\circ\frac{1}{\bnu^\rho}\cdot d\bxi^\rho_t,
 \end{align}
 which are equivalent to the those presented in Sec.~\ref{sec:LNAdyn} and the approximate entropy flow in Sec.~\ref{sec:LNAapprox}.
 
 To summarize the calculations of this Appendix, we have seen that when carrying out the system size expansion on the joint populations and entropy-flow distribution, the consistency of the expansion forces us to consider only near equilibrium fluctuations.
 From this fact we can conclude that the LNA can only inherent a consistent stochastic thermodynamics at equilibrium.

 \end{widetext}

% If in two-column mode, this environment will change to single-column format so that long equations can be displayed. 
%\begin{widetext}
%$$\mbox{put long equation here}$$
%\end{widetext}

% Figures should be put into the text as floats. 
% Use the graphics or graphicx packages (distributed with LaTeX2e).
% Put the label that you will use with \ref{} command in the braces of the \label{} command.
%
% \begin{figure}
% \end{figure}

% \begin{table}
% \end{tabular}
% \end{table}


\bibliography{PhysicsTexts.bib,FluctuationTheory.bib}

\end{document}